\documentclass[twocolumn]{article}

\usepackage{graphicx}

\newcommand{\be}{\begin{equation}}
\newcommand{\ee}{\end{equation}}
\renewcommand{\k}[2]{\frac{#1}{#2}}
\newcommand{\pab}[2]{\frac{\p #1}{\p #2}}
\newcommand{\pcd}[2]{\frac{\p^2 #1}{\p #2^2}}

\def\p{\partial}
\def\s{\,\,\,\,}
\def\0{^{(0)}}
\def\1{^{(1)}}

\def\={\approx}

\def\z{\zeta}
\def\v{\lambda}

\def\ra{\rightarrow}

\title{A Mathematical Model for Voigt Poro-Visco-Plastic Deformation}
\author{Xin-She Yang \\
{\small Department of Mechanical Engineering,
University of Wales Swansea, Singleton Park, Swansea SA2 8PP, UK}}
\date{}
\begin{document}
\maketitle

\begin{abstract}
A mathematical model for poro-visco-plastic compaction and  pressure solution in porous
sediments has been formulated using the Voigt-type rheological
constitutive relation as derived from experimental data. The governing
equations reduce to a nonlinear hyperbolic heat conduction equation in the
case of slow deformation where permeability is relatively high
and the pore fluid pressure is nearly hydrostatic, while travelling wave
exists in the opposite limit where over-pressuring occurs and the pore
fluid pressure is almost  quasi-lithostatic.
Full numerical simulation using a finite element method agree well
with the approximate analytical solutions. \\
\end{abstract}

\noindent {\bf Citation detail:} X. S. Yang, A mathematical model for Voigt poro-visco-plastic deformation,
{\it Geophys. Res. Lett.}, {\bf 29}(5), 10.1029/2001GL014014 (2002).

\section{Introduction}

Many physical properties such permeability, viscosity, Young's modulus and
thermal conductivity vary with porosity or fluid content in sedements and
minerals. The porosity in turn depends on the deformation and compaction
state and can be calculated from the compaction curves resulting from the
proper compaction modelling [Audet and Fowler, 1992]. Furthmore, compaction
is also related to
overpressuring, mineral deposition, hydrocarbon generation and oil migration
in reservoir. Thus the correct modelling of compaction is both of scientific
importance as well as industrial interest. However, due to the nonlinear
feature in the compaction process and difficulty in formulating the accurate
and realistic rheological relationship in sediments and rocks, most of existing
models use simplified rheology such as poroelastic, purely viscous
or viscoelastic  relations
[{\it Rutter}, 1976; {\it Wangen}, 1992; {\it Holzbercher}, 1998;
{\it Revil}, 1999; {\it Yang},2000]. The rheological properties of
realistic granular sediments are usually viscoelastic or viscoplastic as
implied by experiments. {\it  Yang} [2000] presents a viscoelastic model
of Maxwell type and comparison of analytical solutions
with the numerical simulations shows very good
agreement. However, {\it Revil}'s [1999] work suggests that it maybe more
appropriate to use a poro-visco-plastic model of Voigt type.
We intend to
do the analysis similar to {\it Yang}'s [2000] but using the
Voigt-type consitutive relation as derived by {\it Revil} [1999].

Much of the work in this area has been reviewed by {\it Rieke and Chilingarian}
[1974], {\it Birchwood and Turcotte} [1994] and later by  {\it Fowler and Yang}[1999]. 
This paper aims at providing  a new approach to compaction
and pressure solution by using {\it Revil}'s visco-poro-plastic
relation of Voigt type [{\it Revil}, 1999]. The nonlinear partial
differential equations are then analysed by using asymptotic
methods and the obtained analytical solutions are compared with numerical
simulations. Although the present work mainly concerns  the 1-D theorectical
formulation and analytical solution procedure, however, we intend to provide a
simplified and yet realistic framework for further research in this area
and shows how compaction mechanism is related to rheological relationships
and material properties of porous sediments, so that more realistic
constitutive relationships can be formulated and analysed.

\section{MATHEMATICAL MODEL}

The fundamental model of compaction and pressure solution is essentially
similar to the model of soil consolidation process.
The solid sediments act as a
compressible porous matrix, so mass conservation of pore °fluid together
with Darcy's law leads to an equation of the general type.
Based on earlier work by {\it Audet and Fowler} [1992], {\it Revil}[1999]
and {\it Yang}[2000],
we can  write down the poro-visco-plastic compaction model  of  Voigt type.
In the one-dimensional  case, we have the following governing equations
\begin{equation}
\pab{ (1-\phi)}{t} +\pab{}{z}[ (1-\phi) u^{s} ]=0,
\label{VC:MASS}
\end{equation}
\begin{equation}
\pab{ \phi}{t}+ \pab{( \phi  u^{l})}{z}=0,
\end{equation}
\be
\phi (u^{l}-u^{s})
=-\frac{k(\phi)}{\mu}[ \pab{p}{z}+\rho_l g ],
\end{equation}
\be
- G \pab{p_e}{z}-\pab{p}{z}-\rho g=0, \s
\rho=\rho_s (1-\phi)+\rho_l \phi,
\ee
where $\phi$ is porosity.
 $u^{l} $ and $u^{s}$ are the velocities of fluid and solid matrix,
respectively.
$k$ and $\mu$ are the matrix  permeability and the liquid viscosity,
$\rho_{l} $ and $\rho_{s}$ are the densities of fluid and solid matrix,
$p_{e}$ is the effective pressure,  $p^{l}$ is the pore pressure,
and $g$ is the gravitational acceleration.
$G=1+4 \eta_0/3 \xi_0$ is a constant describing the material properties
with $\eta_0$ and $\xi_0$ being the shear modulus and bulk viscosity
[{\it Bird et al}, 1977].
The first two equations are the conservation of mass for the solid phase and
liquid phase, respectively. The third equation is the Darcy's law in 1-D form
and the last equation is actually the force balance in a simplified form whose
detailed derivation can be found in [{\it Fowler and Yang}, 1998].
Combining equation (3) and (4), we have
\be
\phi (u^{l}-u^{s})
=\frac{k(\phi)}{\mu}[-G \pab{p_e}{z}-(\rho_s-\rho_l)(1-\phi) g ],
\ee
In writing the above equations, we have used an upward coordinate $z$ originating from $z=0$, 
which corresponds to the bottom of the sedimentary column, so that the
ocean floor $z=h(t)$ moves as compaction proceeds. We use such a coordinate
system because it simplifies the analytical solution procedure and also in
keeping with the similar lines of earlier work in
this area [{\it Audet and Fowler}, 1992; {\it Yang}, 2000].
However, the conventional depth coordinate is simply $z-h(t)$, thus
the transformation shall be straightforward once the basin thickness
$h(t)$ is known.  As we shall see in the later
sections, we provide an explicit formula for $h(t)$ as a very good
approximation.

In addition, a rheological  compactional relationship derived from experimental
data [Revil, 1999] is needed to complete this  model in the form
\be
p_e=-\xi \pab{u^s}{z}-E \int^{t}_{0} \pab{u^s}{z} dt,
\ee
where $E$ is the elastic modulus and $\xi$ is the viscosity modulus.
 There are
essentially the same parameters as introduced by {\it Revil} [1999].
The first term of the right hand of the equation is the usual contribution
by viscous plastic deformation, while the second term corresponds to the
poro-elastic deformation.

\section{Non-dimensionalization}

To write the governing equations in dimensionless forms, typical length and
time scales are required. For a typical sedimentation
rate $\dot m_s$, the corresponding typical time scale is $d/\dot m_s$
where the typical length scale $d$ can be defined as
\begin{equation}
d=\{\k{\xi \dot m_s G}{(\rho_s-\rho_l) g} \}^{1/2},
\ee
so that the dimensionless pressure $p=G p_{e}/(\rho_{s}-\rho_{l}) g d=O(1)$.
Meanwhile,  we  scale $z$ with $d$, $u^{s}$ with
${\dot m}_{s}$, time $t$ with $ d/{\dot m}_{s}$, permeability $k$ with
$k_{0}$. By writing $k(\phi)=k_{0} k^*$, $z=d z^*$, ..., and dropping
the asterisks, we thus have
\begin{equation}
-\pab{\phi}{t} +\pab{}{z}[(1-\phi) u^{s}]=0,
\label{VC:MASS-1}
\end{equation}
\begin{equation}
\pab{\phi}{t}+ \pab{(\phi u^{l})}{z}=0, \label{PHI:U}
\end{equation}
\begin{equation}
\phi (u^{l}-u^{s})=\v k(\phi) [-\pab{p}{z}-(1-\phi) ],
\end{equation}
\be
p=-\pab{u^s}{z}-\Xi \int^{t}_{0} \pab{u^s}{z} dt
\ee
where
\begin{equation}
\lambda=\frac{k_{0} (\rho_{s}-\rho_{l}) g} {\mu {\dot m}_{s}},
\s \Xi=\k{E G}{(\rho_s-\rho_l) g d}.
\ee
Adding (\ref{VC:MASS-1}) and (\ref{PHI:U}) together and integrating
from the bottom, we have
\be
u^s=-\phi (u^l-u^s),
\ee
where $u=\phi (u^l-u^s)$ is the Darcy flow velocity.
Now, we  have
\begin{equation}
\pab{\phi}{t} =\pab{}{z}[(1-\phi) u^s],
\label{equ-1}
\end{equation}
\begin{equation}
u^s=\v (\k{\phi}{\phi_0})^m [-\pab{p}{z}-(1-\phi) ].
\label{equ-2}
\end{equation}
\be
p=- \pab{u^s}{z}-\Xi \int^{t}_{0} \pab{u^s}{z} dt,
\label{equ-3}
\ee
where we have used the nonlinear constitutive relation for permeability
$k(\phi)$  of typical form [{\it Smith}, 1971]
\be
k(\phi)=(\k{\phi}{\phi_0})^m,
\ee
where $\phi_0$ is the initial depositional porosity. The exponent $m$
has a typical value of $3 \sim 6$ for sands and sandstones.

The boundary conditions are
\be
\pab{p}{z}-(1-\phi)=0 \s ({\rm or \,\, equivalently,} \s u^s=0),
\s  {\rm at } \s z=0,
\label{bbb-1}
\ee
\[
\phi=\phi_{0}, \,\,\,p=0, \] \be
\dot h=\dot m_s + \v (\k{\phi}{\phi_0})^m [\pab{p}{z}-(1-\phi)]
\s {\rm at} \s z=h(t).
\label{bbb-2}
\end{equation}

It is useful  to estimate these parameters  by using values taken
from observations. By using the typical values
of $\rho_{l} \sim 10^{3} \, {\rm kg\, m}^{-3},\,
\rho_{s} \sim 2.5 \times 10^{3} \, {\rm kg\, m}^{-3},\,$
$ k_{0} \sim 10^{-15} - 10^{-20}\, {\rm m}^{2}, \, \mu \sim 10^{-3}\,
{\rm N\,s\,
m}^{2}, \, \xi  \sim 1 \times 10^{21}$ N s $ {\rm m}^{-2}, $
$\dot m_{s}  \sim 300\, {\rm m\,\, Ma}^{-1}=1 \times 10^{-11}\,
{\rm m\,\, s}^{-1},\, g \sim 10 {\rm m \,s}^{-2}, \, E \sim 10^{9} {\rm N/m}^2,
\, G \sim 1, \, d \sim 1000 {\rm m}$ ;
then $\v \= 0.01 \sim 1000$ and $\Xi \sim 40$. We can see that the main
parameters $\v$ and $\Xi$, which govern the
evolution of the fluid flow and porosity in sedimentary basins,
are the ratios of permeability to sedimentation rate and material
modulus to the typical pressure scale. As $\v$ is essentially controlled
by the hydraulic conductivity, so it becomes the dominant parameter controlling
the whole compaction and pressure solution processes.

\section{Asymptotic Analysis}

Since the nondimensional parameter $\v \=0.01 \sim 1000$ varies greatly and
essentially controls  the  compaction process, we can expect that the two
distinguished limits ($\v \ll 1$ and $\v \gg 1$) will have very different
features in porosity and flow evolutions.  In fact,
$\lambda=1$ defines a transition between slow
compaction ($\lambda \ll 1$) and fast compaction ($\lambda \gg 1$).
The case of $\v \ll 1$ corresponds to the situation where the pore fluid
pressure is nearly hydrostatic whereas the opposite case corresponds to an
overpressured section in which the pore fluid pressure is quasi-lithostatic.
Thus we can follow the similar asymptotic analysis [{\it Fowler and Yang},
1998,1999] to obtain some analytical asymptotic solutions.

\subsection{Slow Deformation ($\v \ll 1$)}

In the nearly hydrostatic case of $\v \ll 1$,
$z \sim 1$, $t \sim 1$, $p \sim 1$ implies that
$u^s \ll 1$ and $\pab{\phi}{t} \= 0$, then $\phi \= \phi_0$.
We thus have
\be
\pab{\phi}{t} \= -\v (1-\phi_0) \pcd{p}{z}, \label{equ-100}
\ee
\be
u^s \=\v [-\pab{p}{z}-(1-\phi_0)],
\ee
and using (\ref{equ-100}), we have
\[
p \=-\pab{u^s}{z}-\Xi \int^{t}_0 \k{1}{(1-\phi_0)} \pab{\phi}{t} dt \] 
\be =-\k{1}{(1-\phi_0)} (\pab{\phi}{t}+\Xi \phi),
\ee
combining these above three equations, we have a single equation
for $\phi$
\be
\pab{\phi}{t}=\v \Xi \pcd{\phi}{z}+\v \k{\p^3 \phi}{\p t \p z^2}.
\ee
As the $\Xi \gg 1$ and $\v \ll 1$, we can use the approximation
$\phi_t \=\v \Xi \phi_{zz}$ so that we have
\be
\pab{\phi}{t}=\v \Xi \pcd{\phi}{z}+\k{1}{\Xi} \pcd{\phi}{t}.
\ee
with appropriate boundary conditions
\be
\pab{\phi}{z} \= -\k{(1-\phi_0)^2}{\Xi}, \s {\rm on} \s z=0,
\ee
\be
\phi \ra \phi_0, \s z \ra \infty,
\ee
This problem is in fact equivalent to the problem of hyperbolic heat
conduction or non-Fourier heat equation
 which is well-documented
in heat transfer and laser pulse modelling [{\it Antaki}, 1997].
By using the Laplace transform
method,  we can write the solution approximately in terms of Bessel
functions $J_i$ as
\[
\phi \=(1-\phi_{0}) \sqrt{4 \lambda \Xi t}\,\,
{\rm ierfc}(\z) \] \be -\k{(1-\phi_0)^2}{\Xi} 
[ \sqrt{\v \Xi t}+\k{\v}{4 \Xi} \sum_{i=1}^{\infty} \k{J_0(z
\alpha_i)}{J_0(\alpha_i) \alpha_i} ],
\end{equation}
where
\begin{equation}
{\rm ierfc}(\z)=\frac{1}{\sqrt{\pi}} e^{-\z^{2}}-\z {\rm erfc}(\z).
\end{equation}
and $\alpha_i$ is the {\it i}th real non-negative  root
of equation $J_1(\alpha_i)=0$.
We can see that compaction essentially occurs in a boundary layer
near the bottom with a thickness of the order of $\sqrt{\v}$.

\subsection{Fast Deformation ($\v \gg 1$)}

In the case of $\v \gg 1$, the dependence of permeability on porosity
$(\phi/\phi_0)^m$ decrease dramatically, so that $\v (\phi/\phi_0)^m$ is
only bigger enough when $\phi>\phi_*=\phi_0 \exp[-(\ln \v)/m]$.
Thus, we have
\be
\pab{\phi}{t} \= (1-\phi_*) \pab{u^s}{z}, \label{equ-200}
\ee
\be
\pab{p}{z} \= (1-\phi),
\ee
and using the equation (\ref{equ-200}),   we get
\[
p=-\k{1}{(1-\phi_*)} \pab{\phi}{t}-\Xi \int^{t}_0 \k{1}{(1-\phi_*)}
\pab{\phi}{t} dt \] \be =  -\k{1}{(1-\phi_*)} (\pab{\phi}{t} +\Xi \phi).
\ee
Combining these equation, we can get a single equation for $\phi$
\be
(1-\phi)(1-\phi_*) =\k{\p^2 \phi}{\p t \p z}+\Xi \pab{\phi}{z}.
\ee
Now we can seek the traveling wave solution of the form $\phi=\phi(\z)$
with $\z=z-c t$, so that we have
\be
(1-\phi)(1-\phi_*) =-c \phi''+\Xi \phi'
\ee
where $\phi'=d \phi/d \z$. We can easily write the solution as
\be
\phi=1-(1-\phi_0) \exp[\k{[\Xi-\sqrt{\Xi+4 c (1-\phi_*)}] \z}{2 c}]
\ee
In fact, the above solution is only valid for the top  part when $\phi <
\phi_*$. Using equation (6) and $\phi \sim \phi_*$ as $\z \ra -\infty$,
the travelling wave implies that \be
c \phi+ (1-\phi) u^s= c \phi_0 + (\dot m_s-c) (1-\phi_0).
\ee
so that we have
\be
c \= \dot m_s (\k{1-\phi_0}{1-\phi_*}), \s h(t) \=\dot m_s
(\k{1-\phi_0}{1-\phi_*} t,
\ee
which means the basin thickness increases linearly with time.

\section{Numerical Simulations}

In order to check the accuracy of the above analysis, we used
a finite element method to solve the above equations (8)-(11).
For simplicity, we only present the related results in Fig.1
where $t=5$ with different values of the $\v=0.001, 0.1, 10, 1000$,
$\phi_0=0.5$ and $\Xi=40$. This good agreement near the top and
the bottom regions suggest that there exists a travelling wave solution
on the top for $\v \gg 1$ and a boundary layer near the bottom for
the small $\v$ case.

\section{Discussion}

A mathematical model for poro-visco-plastic comapction and pressure solution in
porous sediments has been formulated using the Voigt-type rheological
constitutive
relation as derived from experimental data by {\it Revil} [1999].
After the proper scalings,
the governing equations reduce to a system of coupled partial differential
equations of mixed type. In the case of small $\v$ where the sedimentation
is fast, permeability is small and the pore fluid pressure is nearly
hydrostatic, the pressure solution process reduces to the case of
hyperbolic heat conduction equation with a boundary layer forming at
the bottom of the compacting column.  On the other hand, for the large
$\v \gg 1$ case where sedimentation or loading is slow in  high permeable
sediments, the travelling wave solution exists and the top surface
moves up with nearly constant velocity.

Compared with the earlier work [{\it Yang}, 2000]
using the viscoelastic rheological
relation, we see that the there is no essential difference in the case
small deformation ($\v \ll 1$). Boundary layer exists in both viscoelastic and
poro-viscous-plastic cases although the  slight difference is that the
former viscoelastic case corresponds to the heat conduction with a constant
flux and a constant source term, while the latter poro-visco-plastic case
is mainly a hyperbolic heat condution mechanism. The case of fast deformation
and compaction ($\v \gg 1$) is more complicated.
Although both viscoelastic and poro-viscous-plastic
cases have a transition at the depth where $\phi \=\phi_*$, however,
the mechanisms
above and below the transition are very different.
In the viscoelastic case, the
top region is essentially poroelastic while the lower region is almost purely
viscous, while in the present poro-visco-plastic case, the
mechanism is a complicated combination of
viscous-plastic mechanism and poro-elastic deformation process
controlled by the
proper balance of the parameters $\v$ and $\Xi$.  Furthermore, in comparison
with the earlier results, this work suggests that Voigt-type poro-visco-plastic
deformation has much interesting characteristics due to its
time-dependence feature. \\

{\bf Acknowledgments:} The author  would like to thank the referee(s) for their instructive and helpful
comments which have greatly improved the original manuscript.

\begin{figure}
\centerline{\includegraphics[width=3in]{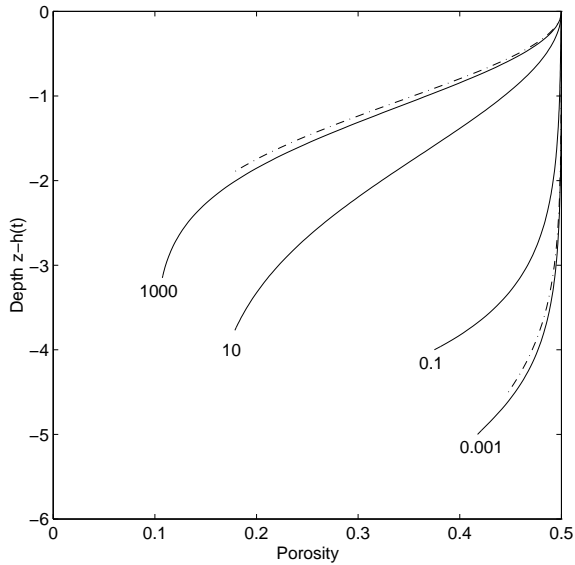}}
\caption{Comparison of numerical simulations (solid) with analytical solutions (dashed curves)
for different values of $\lambda=0.001 \sim 1000$ as marked on the curves. }

\end{figure}

\end{document}